\documentclass[twocolumn,prd,aps,longbibliography,superscriptaddress,preprintnumbers,tightenlines,
showpacs,nofootinbib,eqsecnum,amsfonts,amsmath]{revtex4}
\pdfoutput=1

\usepackage{color}
\usepackage{calc}
\usepackage{graphicx}
\usepackage{tensor}
\usepackage{bm}
\usepackage{url}
\usepackage{times}
\usepackage{multirow}
\usepackage[varg]{txfonts}
\usepackage{float}
\usepackage{dcolumn}
\usepackage[nolist,nohyperlinks]{acronym}
\usepackage{xspace}
\usepackage[english]{babel}
\usepackage[abs]{overpic}
\usepackage{pict2e}
\usepackage[caption=false]{subfig} % prevent loading of caption package: http://tex.stackexchange.com/questions/22388/subfigures-with-revtex
\allowdisplaybreaks[1]
\usepackage[utf8]{inputenc}
\usepackage{braket}
\usepackage{stackengine}
\usepackage{gensymb}
\usepackage[dvipsnames]{xcolor}

\usepackage[colorlinks]{hyperref}
\usepackage{boldline}
\usepackage{longtable}

\usepackage{soul}
\DeclareMathAlphabet{\mathcalstd}{OMS}{cmsy}{m}{n}
\DeclareMathAlphabet{\mathpzc}{OT1}{pzc}{m}{it}

\newcommand{\UIB}{Departament de F\'isica, Universitat de les Illes Balears, IAC3 -- IEEC, Crta. Valldemossa km 7.5, E-07122 Palma, Spain}
\newcommand{\Zurich}{Physik-Institut, Universität Zürich, Winterthurerstrasse 190, 8057 Zürich, Switzerland}

\begin{document}
\preprint{LIGO-P2000179}

%%%%%%%%%%%%%%%%%%%%%%%%%%%%%%%%%%%%%%%%%%%%%%%%%%%% Title page %%%%%%%%%%%%%%%%%%%%%%%%%%%%%%%%%%%%%%%%%%%%%%%%%%%%

\title{Impact of eccentricity on the gravitational wave searches for binary black holes: High mass case}

\author{Antoni Ramos-Buades}
\affiliation{\UIB}
\author{Shubhanshu Tiwari}	
\affiliation{\Zurich}
\author{Maria Haney}
\affiliation{\Zurich}
\author{Sascha Husa}
\affiliation{\UIB}
%\author{Order to be decided, and other people?}
%\author{Geraint Pratten}
%\affiliation{\UIB}
%\affiliation{\UoB}
%\author{Cecilio García Quirós}
%\affiliation{\UIB}
%\author{Marta Colleoni}
%\affiliation{\UIB}
%\author{Héctor Estellés}
%\affiliation{\UIB}
%\author{Maite Mateu}
%\affiliation{\UIB}
%\author{Rafel Jaume}
%\affiliation{\UIB}

%\affiliation{\ICTS}

%%%%%%%%%%%%%%%%%%%%%%%%%%%%%%%%%%%%%%%%%%%%%%%%%%% Abstract %%%%%%%%%%%%%%%%%%%%%%%%%%%%%%%%%%%%%%%%%%%%%%%%%%%%%%
\begin{abstract}
The possible formation of stellar-mass binary black holes through dynamical interactions in dense stellar environments predicts the existence of binaries with non-negligible eccentricity in the frequency band of ground-based gravitational wave detectors; the detection of binary black hole mergers with measurable orbital eccentricity would validate the existence of this formation channel.
Waveform templates currently used in the matched-filter gravitational-wave searches of LIGO-Virgo data neglect effects of eccentricity which is expected to reduce their efficiency to detect eccentric binary black holes. Meanwhile, the sensitivity of coherent unmodeled gravitational-wave searches (with minimal assumptions about the signal model) have been shown to be largely unaffected by the presence of even sizable orbital eccentricity. In this paper, we compare the performance of two state-of-the-art search algorithms recently used by LIGO and Virgo to search for binary black holes in the second Observing Run (O2), quantifying their search sensitivity by injecting numerical-relativity simulations of inspiral-merger-ringdown eccentric waveforms into O2 LIGO data. Our results show that the matched-filter search PyCBC performs better than the unmodeled search cWB for the high chirp mass ($>20 M_{\odot}$) and low eccentricity region ($e_{30 Hz} < 0.3$) of parameter space. For moderate eccentricities and low chirp mass, on the other hand, the unmodeled search is more sensitive than the modeled search.  
\end{abstract}

\pacs{
04.25.Dg, % Numerical studies of black holes and black-hole binaries
04.25.Nx, % Post-Newtonian approximation; perturbation theory; related approximations
04.30.Db, % GW Wave generation and sources
04.30.Tv  % GW Gravitational-wave astrophysics
}

\today

\maketitle

% ======================
%  ACRONYMS
% ======================
\acrodef{PN}{post-Newtonian}
\acrodef{EOB}{effective-one-body}
\acrodef{NR}{numerical relativity}
\acrodef{GW}{gravitational-wave}
\acrodef{BBH}{binary black hole}
\acrodef{BH}{black hole}
\acrodef{BNS}{binary neutron star}
\acrodef{NSBH}{neutron star-black hole}
\acrodef{SNR}{signal-to-noise ratio}
\acrodef{aLIGO}{Advanced LIGO}
\acrodef{AdV}{Advanced Virgo}

\newcommand{\PN}[0]{\ac{PN}\xspace}
\newcommand{\EOB}[0]{\ac{EOB}\xspace}
\newcommand{\NR}[0]{\ac{NR}\xspace}
\newcommand{\BBH}[0]{\ac{BBH}\xspace}
\newcommand{\BH}[0]{\ac{BH}\xspace}
\newcommand{\BNS}[0]{\ac{BNS}\xspace}
\newcommand{\NSBH}[0]{\ac{NSBH}\xspace}
\newcommand{\GW}[0]{\ac{GW}\xspace}
\newcommand{\SNR}[0]{\ac{SNR}\xspace}
\newcommand{\aLIGO}[0]{\ac{aLIGO}\xspace}
\newcommand{\AdV}[0]{\ac{AdV}\xspace}

%%%%%%%%%%%%%%%%%%%%%%%%%%%%%%%%%%%%%%%%%%%%%%%%%%%%%%%
\section{Introduction}\label{sec:introduction}
The number of detections of gravitational wave (GW) signals has steeply increased from the first and second Observing runs (O1/O2) of Advanced LIGO and Advanced Virgo \cite{LIGOScientific:2018mvr} to the third Observing run (O3), where tens of GW candidates have already been recorded~\cite{GraceDBO3,Abbott:2020uma,LIGOScientific:2020stg,GWTC2}. So far, all GW detections of binary black holes (BBHs) are consistent with signals emitted from quasicircular binaries \cite{Salemi:2019owp, Romero-Shaw:2019itr}.

Generally, two main scenarios can be considered regarding possible formation channels for BBH mergers: 1) isolated binary evolution \cite{Bethe:1998bn,Belczynski_2002,10.1093/mnras/sty2190,Barrett:2017fcw}, during which BBHs shed their formation eccentricity through GW emission and have circularized by the time they enter the frequency band of the ground-based detectors \cite{PhysRev.131.435,PhysRev.136.B1224}; 2) binaries dynamically formed in dense stellar environments like globular clusters and active galactic nuclei~\cite{O_Leary_2006,PhysRevD.97.103014,Fragione:2019hqt, Kumamoto:2018gdg, OLeary:2008myb}, which may still retain a significant eccentricity by the time they enter the frequency band of the Advanced LIGO~\cite{TheLIGOScientific:2014jea} and Advanced Virgo~\cite{TheVirgo:2014hva} detectors. Although both formation channels (and their different astrophysical scenarios) predict BBH mergers with distinct distributions of masses and spins \cite{LIGOScientific:2018jsj,Mandel:2009nx,TheLIGOScientific:2016htt,Farr:2017uvj}, the model uncertainties \textemdash as well as the low statistics due to the limited number of GW detections \textemdash do not permit to set tight constraints on BBH formation scenarios from the mass and spin distributions alone.

Dynamical BBH formation, however, is distinctly characterized by the potential existence of binaries with non-negligible eccentricity in the frequency band of the ground-based detectors, which were formed through dynamical capture at very close separations (without time to circularize before merger) or through a dynamical process that increased the eccentricity of the binary (e.g. Kozai-Lidov oscillations \cite{1962P&SS....9..719L,1962AJ.....67..591K}). The detection of a GW signal with an unambiguous signature of non-negligible orbital eccentricity would therefore confirm the dynamical formation channel for BBHs and provide information about possible formation mechanisms and the astrophysical environments of such sources.

In order to be able to confidently detect eccentric binary black hole signals it is necessary to assess the sensitivity of the pipelines used to search for such signals. As a consequence several studies have analysed the sensitivity of different search pipelines to eccentric compact binary mergers over data from O1 and O2 Advanced LIGO and Advanced Virgo observing runs \cite{V_ebbh,ebbho1o2,Nitz:2019spj}.

In this paper we quantify the sensitivity of two different gravitational-wave search pipelines to eccentric inspiral-merger-ringdown (IMR) signals calculated from numerical relativity (NR) simulations. The two search pipelines are: 1) the template-based PyCBC algorithm \cite{PhysRevD.85.122006,Usman:2015kfa}, and 2) the unmodeled coherent WaveBurst (cWB) algorithm \cite{Klimenko:2008fu,PhysRevD.93.042004}. We study the sensitivity of the pipelines with increasing eccentricity of the signal for three different mass ratios $q=1,2,4$, with $q=m_1/m_2>1$ and $m_1$, $m_2$ the component masses of the binary. Furthermore, for mass ratio $q=1$ we inject eccentric simulations with increasing dimensionless component spins $|\vec{\chi}_i| \leq 0.75$ (aligned with the orbital angular momentum of the system), where $\vec{\chi}_i= \vec{S}_i/m^2_i$ and $\vec{S}_i$ the spin vector of the i-component, with $i=1,2$. Due to the restricted length of the NR simulations the waveforms are injected at a start frequency of $30$Hz, and the eccentricity is consistently defined at that frequency according to the procedure detailed in Sec. ~\ref{sec:introEcc}.

%\toni{ We need a short paragraph here with an overview of the results for both pipelines?}

The paper is organised as follows: In Sec.~\ref{sec:introEcc} we provide details about the IMR NR eccentric waveforms used in this work. In Sec. \ref{sec:introSearches} we briefly summarize the two search algorithms considered in this study, the template-based search PyCBC and the un-modeled search, cWB. We present in Sec. \ref{sec:introIFAR} the results of the sensitivity estimates of both studied pipelines. We conclude in Sec. \ref{sec:summary} discussing the results obtained and reporting our conclusions. 
 
 %%%%%%%%%%%%%%%%%%%%%%%%%%%%%%%%%%%%%%%%%%%%%%%%%%%%%%%%
\section{Eccentric binary black holes} \label{sec:introEcc}
The gravitational wave signals emitted from generic binary black holes are described by 17 parameters \cite{Sathyaprakash:2009xs}. The parameters of a binary can be separated into  10 intrinsic parameters, i.e. properties of the emitting source, and 7 extrinsic parameters, describing the position of the source in the detector sky. The intrinsic parameters are the two component masses $m_i$, the six dimensionless spin vectors $\vec{\chi}_i=\vec{S}_i/m^2_i$, the eccentricity parameter $e$, and the argument of the periapsis  $\Omega$.  Another useful mass parameter in gravitational wave data analysis is the chirp mass $\mathcal{M}$ of a binary with masses $m_1$ and $m_2$, which is defined as $\mathcal{M}\equiv (m_{1}m_{2})^{3/5}(m_1+m_2)^{-1/5}$.

The extrinsic parameters are the luminosity distance $d_L$, the azimuthal angle $\varphi$, the inclination $\iota$, the time of coalescence $t_c$, the polarization angle $\psi$, the right ascension $\phi$ and the declination $\theta$. The strain induced in a gravitational wave detector can be written in terms of these parameters as \cite{PhysRevD.47.2198,PhysRevD.58.063001}
\begin{equation}
\begin{split}
h(t,\zeta,\Theta) & =   F_+      (\theta,\phi,\psi) \, h_+(t-t_c;\iota,\varphi, \zeta)  \\
                  & +   F_\times (\theta,\phi,\psi) \, h_\times (t-t_c;\iota,\varphi, \zeta) ,
\end{split}
\label{eq:eq1}
\end{equation}
where $F_+$, $F_\times$ are the antenna pattern functions, and $\Theta=\{ t_c, r, \theta, \varphi,\psi,\iota, \varphi \}$ and $\zeta=\{m_1, m_2, \vec{S}_1, \vec{S}_2, e, \Omega \}$ represent the sets of extrinsic and intrinsic parameters, respectively. The gravitational wave polarizations $(h_+,h_\times)$ appearing in the detector response can be expressed as a complex waveform strain
\begin{equation}
h(t)=h_+ - i h_\times = \sum_{l=2}^{\infty}\sum_{m=-l}^{l} Y^{-2}_{lm}(\iota, \varphi ) h_{lm} (t-t_c;\zeta),
\label{eq:eq2}
\end{equation}
where  $h_{lm}$ are the $(l,m)$ waveform modes and $Y^{-2}_{lm}(\iota, \varphi )$ the spherical harmonics of spin-weight $-2$. 

\subsection{Numerical Relativity data set}
In this work we inject eccentric NR waveforms produced with the open-source  EinsteinToolkit (ET) code \cite{Loffler:2011ay,maria_babiuc_hamilton_2019_3522086} and the SpEc code \cite{SpEC}. The ET waveforms were presented in  \cite{Ramos-Buades:2019uvh}, and the  SXS ones in~\cite{PhysRevD.98.044015}. The injected waveforms are displayed in Table \ref{tab:tabNRinj}, where we show for each simulation its identifier (ID, an integer number), the simulation name, mass ratio, z-components of the dimensionless spin vectors $(\chi_{1,z},\chi_{2,z})$ and the initial eccentricity measured with the method developed in \cite{Ramos-Buades:2019uvh}.

\begin{table}[h!]
\begin{center}
\resizebox{8.5cm}{!}{
 \def\arraystretch{1.3 }
\begin{tabular}{c c c c c c c c c c c c  c c }
\hline  
\hline
ID &  & Simulation &  &    q  & &${\chi}_{1,z}$ & & $\chi_{2,z}$ &    &  $e^{\text{NR}}_0 $  \\
%\hline
\hline
1&   & \texttt{SXS:BBH:1356}&  &     1.& & $0.$ & & $0.$ & &      $0.09$ \\
%\hline
2&   & \texttt{SXS:BBH:1360}&  &    1.&  &$0.$ &  &$0.$  & &     $0.15$ \\
%\hline
3&   & \texttt{SXS:BBH:1363}&  &   1.&  &$0.$ & & $0.$   & &      $0.23 $ \\
%\hline
4 &   &  \texttt{Eccq1.\_\_0.\_\_0.\_\_et0.5\_D27} &  &   1.&  &$0.$ & & $0.$    	  & &   $0.30$ \\
 \hline
5  &   &  \texttt{Eccq1.\_\_-0.25\_\_-0.25\_\_et0.1\_D14} &  &   1.&  &$-0.25$ & & $-0.25$  & &   $0.07$ \\
6 &   &  \texttt{Eccq1.\_\_-0.5\_\_-0.5\_\_et0.1\_D13} &  &   1.&  &$-0.5$ & & $-0.5$     	  & &   $0.07$ \\
7 &   &  \texttt{Eccq1.\_\_-0.75\_\_-0.75\_\_et0.1\_D13} &  &   1.&  &$-0.75$ & & $-0.75$      	  & &   $0.08$ \\
  \hline
8 &   &  \texttt{Eccq1.\_\_0.25\_\_0.25\_\_et0.2\_D16} &  &   1.&  &$0.25$ & & $0.25$     	  & &   $0.12$ \\
9 &   &  \texttt{Eccq1.\_\_0.5\_\_0.5\_\_et0.2\_D15} &  &   1.&  &$0.5$ & & $0.5$     	  & &   $0.12$ \\
10 &   &  \texttt{Eccq1.\_\_0.75\_\_0.75\_\_et0.2\_D15} &  &   1.&  &$0.75$ & & $0.75$     	  & &   $0.12$ \\
  \hline
11&   & \texttt{SXS:BBH:1365}&  &   2.&  &$0.$ & & $0.$   & &     $0.06$ \\
13 &   &  \texttt{Eccq2.\_\_0.\_\_0.\_\_et0.2\_D16} &  &   2.&  &$0.$ & & $0.$    	  & &   $0.14$ \\
12&   & \texttt{SXS:BBH:1369}&  &   2.&  &$0.$ & & $0.$   & &       $0.20$ \\
14 &   &  \texttt{Eccq2.\_\_0.\_\_0.\_\_et0.5\_D26} &  &   2.&  &$0.$ & & $0.$    	  & &   $0.30$ \\ 
  \hline
15 &   &  \texttt{Eccq4.\_\_0.\_\_0.\_\_et0.1\_D12} &  &   4.&  &$0.$ & & $0.$     	  & &   $0.07$ \\
16 &   &  \texttt{Eccq4.\_\_0.\_\_0.\_\_et0.2\_D15} &  &   4.&  &$0.$ & & $0.$     	  & &   $0.14$ \\ 
17 &   &  \texttt{Eccq4.\_\_0.\_\_0.\_\_et0.5\_D27.5} &  &   4.&  &$0.$ & & $0.$    	  & &   $0.30$ \\ 
\hline
\hline
\end{tabular}
}
\end{center}
\caption{Summary of the injected  NR simulations. The first column denotes the identifier of the simulation, the second column indicates the name of the simulation as presented in \cite{Ramos-Buades:2019uvh, PhysRevD.98.044015}. Next columns show the mass ratio, z-component of the dimensionless spin vectors and the initial NR eccentricity as measured using the procedure detailed in \cite{Ramos-Buades:2019uvh}.}
\label{tab:tabNRinj}
\end{table}

The injected data set is chosen with the following criteria: simulations with IDs $1-4$ are equal mass non-spinning cases which serve as control cases because eccentric equal mass non-spinning binaries have already been studied in the literature \cite{PhysRevD.93.043007}, while simulations with IDs $5-10$ extend the equal mass case to the spinning sector. Finally, simulation sets $11-14$ and $15-17$ allow to test the efficiency of the pipelines at higher mass ratios without including spin effects.

The eccentricity parameter describes the ellipticity of the binary's orbit, values close to 0 indicate a quasi-circular evolution while values close to 1 represent an almost head-on collision. In general relativity the eccentricity is a gauge dependent quantity. As a consequence, a plethora of eccentricity estimators have been developed to measure the eccentricity in numerical relativity simulations \cite{Husa:2007rh,Pfeiffer:2007yz,Tichy:2010qa,Buonanno:2010yk,Mroue:2010re,Puerrer2012,PhysRevD.99.023003}. Eccentricity estimators are combinations of dynamical or wave quantities, like the orbital frequency of the binary, the orbital separation, the gravitational wave frequency of the $(2,2)$ mode, etc.,  measuring the relative oscillations in those quantities due to eccentricity. In this work we measure the eccentricity from the gravitational wave frequency of the $h_{22}$ mode, $\omega_{22}$, following the procedures of \cite{Ramos-Buades:2019uvh}.  We remark that the eccentricities presented in Table \ref{tab:tabNRinj}  are measured from the gravitational wave frequency and their values differ from those presented in \cite{Ramos-Buades:2019uvh} as they were calculated there using the orbital frequency computed from the trajectories of the black holes. 

\begin{figure}[ht!]
\centering
\includegraphics[scale=0.3]{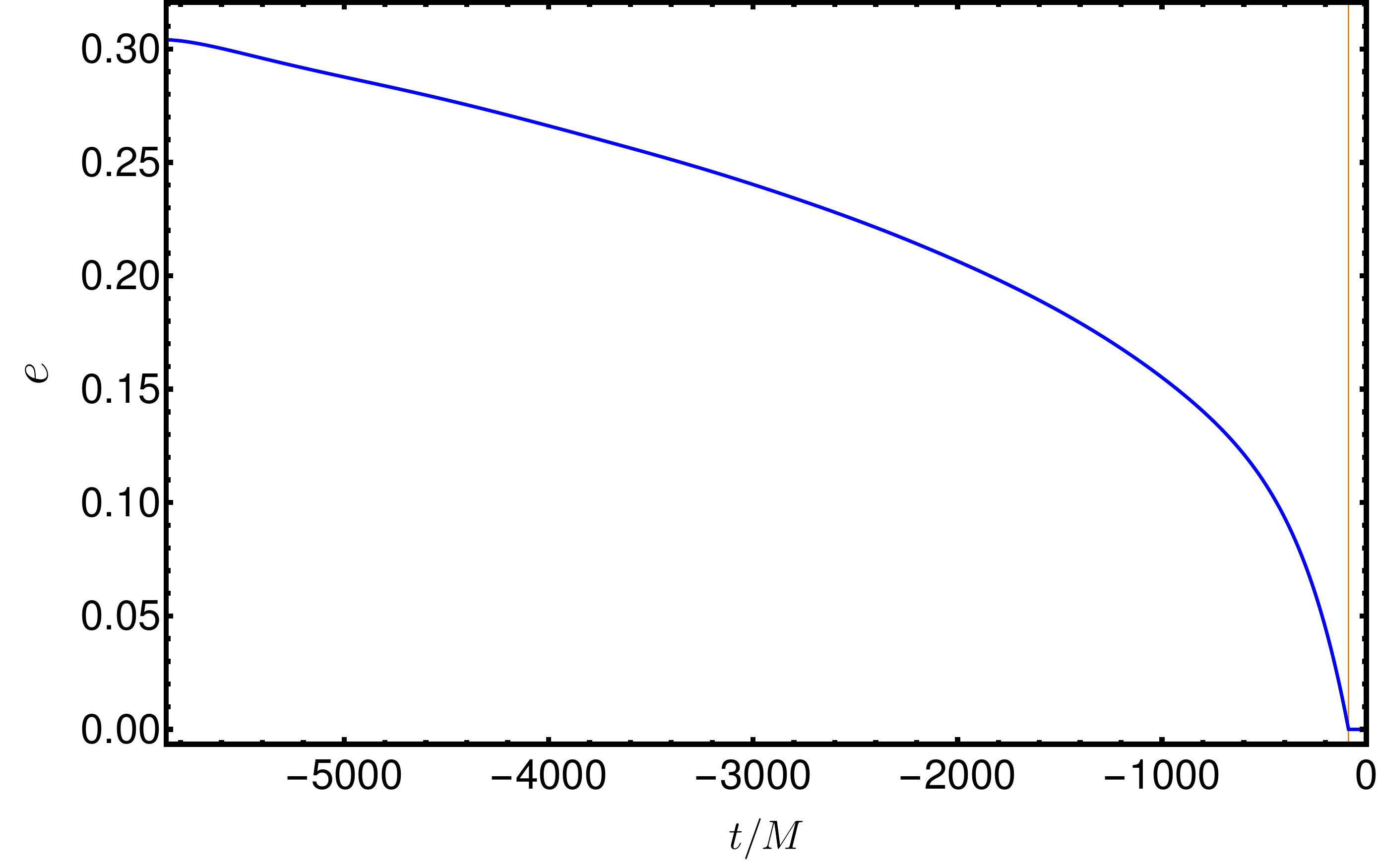}
\includegraphics[scale=0.42]{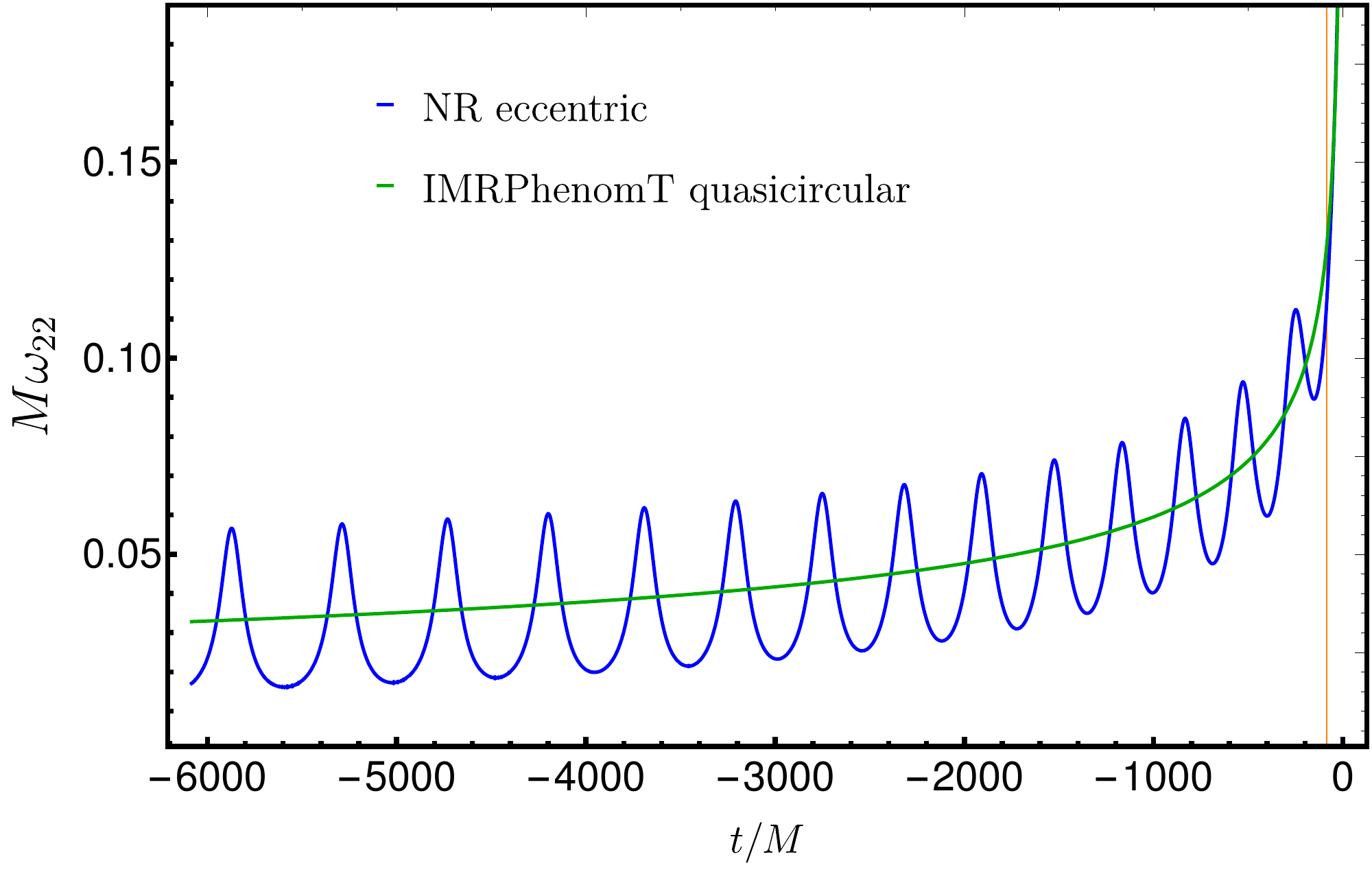}
\includegraphics[scale=0.3]{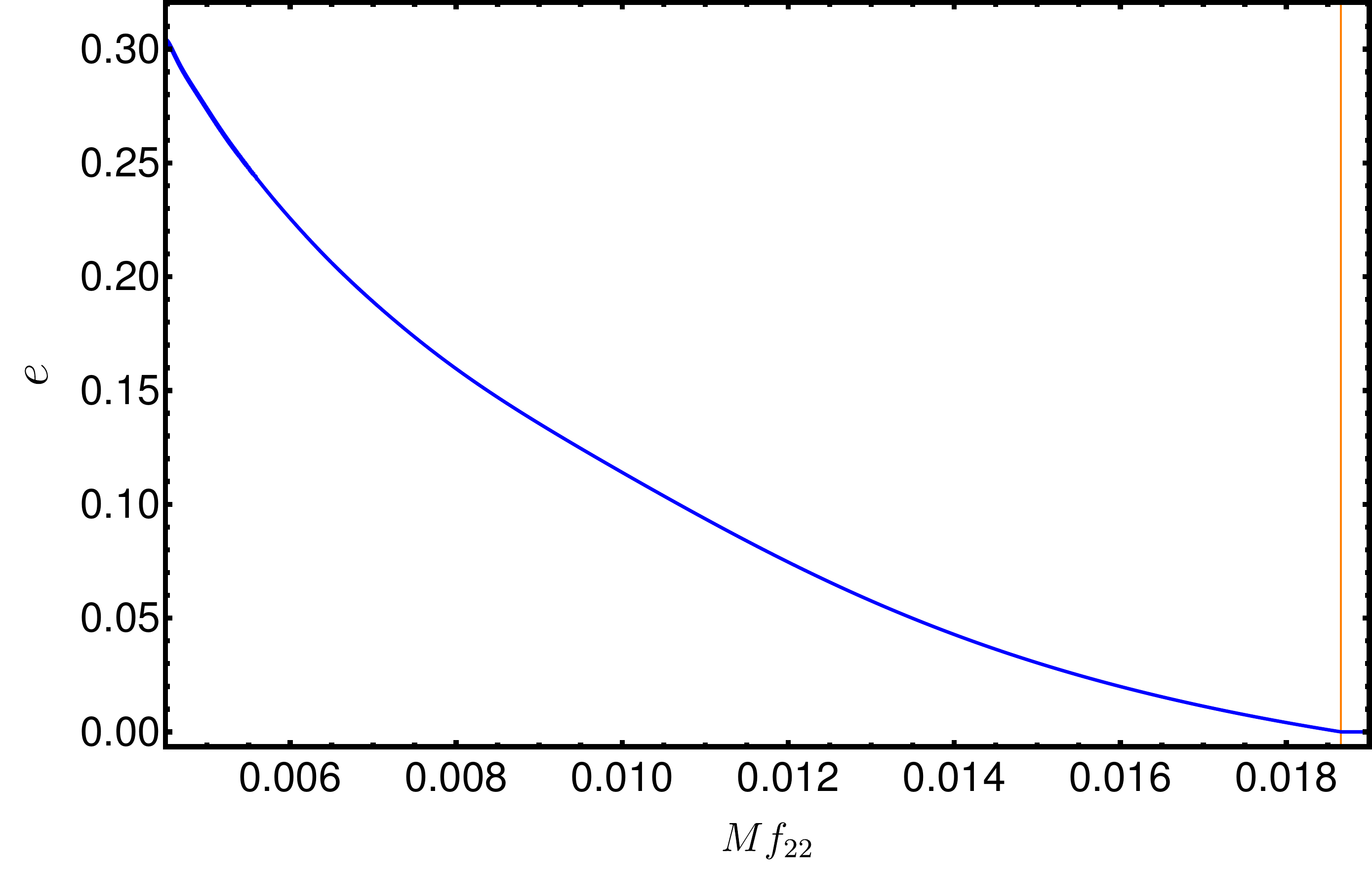}
\caption{Top panel: Time domain evolution of the eccentricity estimated from the eccentric NR simulation with ID 17 in Table \ref{tab:tabNRinj}. Mid panel: Time domain $22$-mode gravitational wave frequencies of the eccentric case with ID 17 from Table \ref{tab:tabNRinj} and of the quasicircular IMRPhenomT waveform model, highlighted in blue and green colors respectively.  Bottom panel: Eccentricity as a function of the gravitational frequency of the $(2,2)$ mode for the same configuration as in the upper panel. With vertical lines in the top and bottom plots we have highlighted the MECO time and frequency, respectively.  }
\label{fig:NRsims}
\end{figure}

In the top panel of Fig. \ref{fig:NRsims} we show the time evolution of the eccentricity of the simulation with ID 17 in Table \ref{tab:tabNRinj}. Moreover, we choose the end of the inspiral given by the minimum energy circular orbit (MECO) \cite{PhysRevD.95.064016}, and explicitly set the eccentricity to zero from the MECO time onwards as at that point the eccentricity is so small which is practically zero. 

In this study we are interested in injecting the waveforms presented in Table \ref{tab:tabNRinj} at a certain detector frequency and for a certain total mass distribution. The modification of the total mass of the system implies a change in length of the waveform within the frequency band of the detector, as a consequence different total masses imply also different initial eccentricities, as one can appreciate from the top panel of Fig. \ref{fig:NRsims}, which shows the eccentricity as a monotonically decaying function as the binary evolves. One possible solution might be to express the eccentricity measured from the NR simulation as a function of gravitational wave frequency of the 22-mode scaled by the total mass of the system, $ 2 \pi M f_{22}=  M \omega_{22} $, approximate the value of the injection frequency by the frequency of the 22-mode, $f_{22} \approx f_{\text{GW}}$, and construct a function $e(M f_{\text{GW}})$  which would provide the value of the eccentricity at a certain total mass for a given injection frequency. However, in the eccentric case the gravitational wave frequency is a non-monotonic function due to the asymmetric gravitational interaction along the orbit of the binary as one can observe in the mid panel of Fig. \ref{fig:NRsims}, where the time domain frequency of the 22-mode for the eccentric simulation with ID 17 from Table \ref{tab:tabNRinj} and the frequency of the quasicircular IMRPhenomT \cite{phenT} waveform model for the same configuration are displayed. We note that after the MECO time both curves converge indicating circularization of the eccentric system at merger.

One possibility for the definition of the eccentricity as a function of a monotonically increasing frequency is to consider the post-Newtonian (PN) approximation, and use the Radiation Reaction (RR) equations \cite{PhysRevD.80.124018} for the PN parameter, x, which can be written in terms of the orbital frequency,  $x= \omega^{2/3}$,  and the temporal eccentricity\footnote{We recall that within the quasi-Keplerian parametrization \cite{1985AIHS...43..107D,Memmesheimer:2004cv} one defines three eccentricities, $e_t$, $e_r$ and $e_\phi$, which can be related to each other by PN expressions. We refer the reader to \cite{PhysRevD.80.124018} for details.} $e_t$. The RR equations are ordinary differential equations for the temporal evolutions of $x$ and $e_t$, derived from the angular momentum and gravitational wave energy fluxes  \cite{PhysRevD.80.124018}. In practice, in the RR equation for $e_t$ one could replace it by the eccentricity measured from the NR simulation and solve the differential equation for $\dot{x}$. However, we find that this procedure does not work satisfactorily, as we have checked that the RR equations show a divergent behavior before the MECO time in some cases, indicating the breakdown of the post-Newtonian approximation. Therefore, we decide to take the gravitational wave frequency of IMRPhenomT and combine it with the eccentricity measured from the simulation to construct the function $e_{NR}(Mf_{22})$. The outcome of such a calculation for the simulation with ID 17 in Table \ref{tab:tabNRinj} is shown in the bottom plot of Fig. \ref{fig:NRsims}. Hence, given an injection with total mass $M_T$ and an injection frequency of $f_{GW}$, we can compute the eccentricity at that frequency and total mass as
\begin{equation}
e_\text{inj}= e_{NR}(M_T f_{GW}).
\label{eq:eq000}
\end{equation}
We note that we focus only on the eccentricity parameter as the initial argument of the periapsis\footnote{Also called initial mean anomaly in the quasi-Keplerian parametrization \cite{1985AIHS...43..107D,Memmesheimer:2004cv}.} in the non-precessing case acts as an initial phase during the inspiral. Its main impact is in the morphology of the waveform at plunge, whose detailed study would require going beyond the maximum total mass   considered in this communication ($M_T>100 M_\odot$). We leave for future work analyzing such high total mass regime.

\begin{figure}
                \includegraphics[scale=0.23]{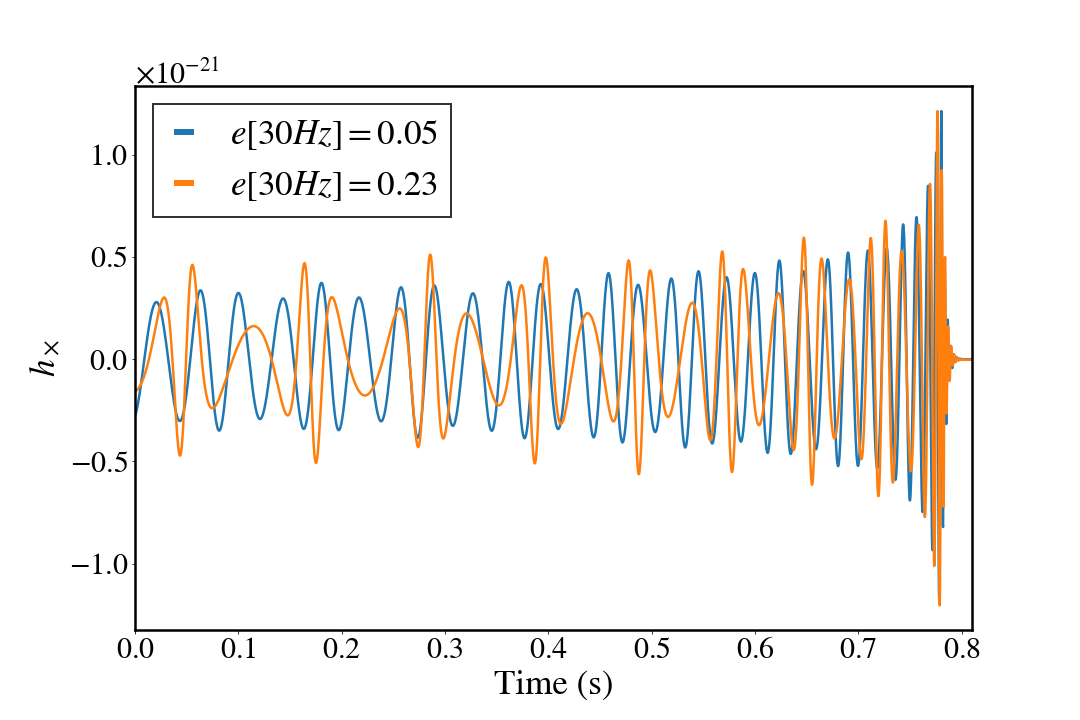}
        \caption{Temporal evolution of the GW polarization state $h_x(t)$ for non-spinning, eccentric stellar-mass binary black holes with total mass $M_T = 50 M_\odot$ and mass ratio $q = 2 $, provided by numerical-relativity simulations. The characteristic orbital eccentricity of the system \textemdash defined at a reference frequency of $30 Hz$ \textemdash is estimated to be $0.05$ (in blue, simulation ID 11) and $0.23$ (in orange, simulation ID 14), respectively.}
        \label{fig:tdwaveforms}
\end{figure}

Finally, in Fig. \ref{fig:tdwaveforms} we plot the time evolution of the GW polarization state $h_x(t)$ for non-spinning, eccentric stellar-mass binary black holes with total mass $M_T = 50 M_\odot$ and mass ratio $q = 2 $, provided by numerical-relativity simulations. The characteristic orbital eccentricity of the system \textemdash defined at a reference frequency of $30 Hz$ \textemdash is estimated to be $0.05$ (in blue, simulation ID 11) and $0.23$ (in orange, simulation ID 14), respectively. The time-domain waveforms clearly demonstrate the effects of increasing initial orbital eccentricity: rapid dephasing, as well as pronounced amplitude modulations due to the advance of the periastron.

\section{Analysis} \label{sec:introSearches}
\subsection{Data} \label{Data} 
The data set used to conduct this study is part of the O2 Data Release through the Gravitational Wave Open Science Center~\cite{Abbott:2019ebz}. This covers approximately $\approx 5$ days of the coincident data between LIGO Livingston and LIGO Hanford between UTC Interval 2017-02-28 16:30:00 - 2017-03-10 13:35:00. Times with significant instrumental disturbances have been removed from the time period considered here \cite{Abbott_2016, Aasi:2012wd}. We consider two search algorithms, PyCBC and cWB, which are described in the following sections. % with the same search configurations that were used for the first catalogue of gravitational waves transients GWTC-1 \cite{LIGOScientific:2018mvr} for these algorithms. 

\subsection{PyCBC : The matched filter algorithm}\label{sec:introMatchFilter}

PyCBC is a search pipeline devised to detect GWs from compact binary coalescences using the PyCBC software package \cite{pycbc1}. In this work we have employed the PyCBC search algorithm in a similar configuration as was used for the first catalogue of gravitational waves transients GWTC-1 \cite{LIGOScientific:2018mvr}. For details of the algorithm see \cite{pycbc1, brucechi2,pycbc3, Nitz_2017,Usman:2015kfa}.

%The analysis uses a template bank of waveforms to perform matched filtering over the data to compute the signal-to-noise ratio (SNR) for each combination of detector, coalescence time and template waveforms \cite{pycbc3}. Triggers are generated by the pipeline according to excesses of matched-filter SNR over a predetermined threshold in each detector. Furthermore,  signal consistency tests between the data and the template, like the $\chi^2$ veto \cite{brucechi2}, are applied to suppress noise transient artefacts (`glitches'). Then, a single-detector rank $\varrho$ is computed for each single-detector trigger using the SNR, the weighting vetoes, and a fitting and smoothing procedure intended to ensure an approximately constant rate of single-detector triggers across the search parameter space~\cite{Nitz_2017}. 
 
 %PyCBC performs also a coincidence test on the remaining triggers \cite{Usman:2015kfa} requiring that the signals observed by the LIGO Hanford and LIGO Livingston detectors have to be seen within a time difference of $12$ ms ($\sim$10 ms travel time between detector $+$2 ms for timing errors). Coincident triggers are assigned a ranking statistic which assesses their statistical significance and approximates the likelihood of obtaining the trigger parameters in the presence of a GW signal versus in the case of only noise~\cite{Nitz_2017}. 

In the PyCBC analysis presented here the template bank described in \cite{75} is used. This bank covers binary systems with a total mass between $2 M_{\odot}$ and $500 M_{\odot}$ and mass ratios $q < 98$. Binary components with masses below  $2 M_{\odot}$ are assumed to be neutron stars with a maximum dimensionless spin magnitude of $0.05$; otherwise the maximum dimensionless spin magnitude is $0.998$. This template bank includes no effects of eccentric orbits.

In a previous study it has been found that a quasicircular bank does not provide a good match for searching binaries with eccentric orbits \cite{brown}. Furthermore, it is known that the signal morphology of the eccentric BBH is orthogonal to the aligned-spin quasicircular BBH \cite{eliu_h}. As a consequence the template bank, which is restricted to the dominant harmonic of quasicircular non-precessing waveforms, becomes ineffective for searching eccentric BBH with high eccentricities.

The way eccentricity affects the matched-filter search by a quasicircular template bank is twofold, \textit{first} the collection of matched filtered signal-to-noise ratio (SNR) is reduced as a function of eccentricity (this can be quantified by studying the overlap of eccentric and quasicircular waveforms), \textit{second} the signal-based $\chi^2$ veto  \cite{brucechi2} used for weighting the single detector SNR to compute the rank also penalizes the final detection statistics of the search. 
 
\subsection{cWB : The un-modeled search algorithm} \label{sec:introcwB}

The cWB  search pipeline  \cite{Klimenko:2008fu,V_ebbh,PhysRevD.93.042004} is designed to detect and reconstruct short-lived signals which are weakly modeled  or unmodeled using a network of GW detectors \cite{PhysRevD.93.042004},  but is also effective for signals with a  known morphology, as is the case of BBH events reported in GWTC-1 \cite{LIGOScientific:2018mvr}.

The configuration of cWB used in this work is the same as used in the GWTC-1 catalog. We refer the reader to \cite{Klimenko:2008fu,V_ebbh,ebbho1o2, PhysRevD.93.042004,chirp_V} for details of the detection process in cWB. The lack of a template bank for binary black holes in eccentric orbits, which could be used by matched filter pipelines, motivated the use of cWB as a robust tool to search  eccentric BBH signals during the first and second observing runs of the LIGO and Virgo detectors \cite{ebbho1o2}.

The cWB search pipeline performs worse than matched-filter pipelines in the case the signal is well recovered by the template bank. In this case matched-filtering would indeed be the optimal method for Gaussian and stationary noise (these simplifications do however not apply to actual LIGO noise).
The sensitivity of cWB  significantly  improves in parts of the parameter space where the template bank does not faithfully reproduce the incoming signal. In an earlier version of cWB, its search sensitivity was found to have almost no dependency on eccentricity \cite{V_ebbh}. This was also confirmed in the latest results for observing runs O1 and O2 of the LIGO and Virgo detectors \cite{ebbho1o2}. As a weakly modelled search cWB is more affected by the background noise and hence has a lower sensitivity as compared to matched filter searches for known signals. Nevertheless cWB has been found to provide a valuable complementarity to matched filter searches to detect signals which are outside the template bank, as in the case of eccentric BBH signals   \cite{V_ebbh} or intermediate mass BBH signals \cite{CalderonBustillo:2017skv}.

\begin{figure*}[!]
  	\includegraphics[scale=0.23]{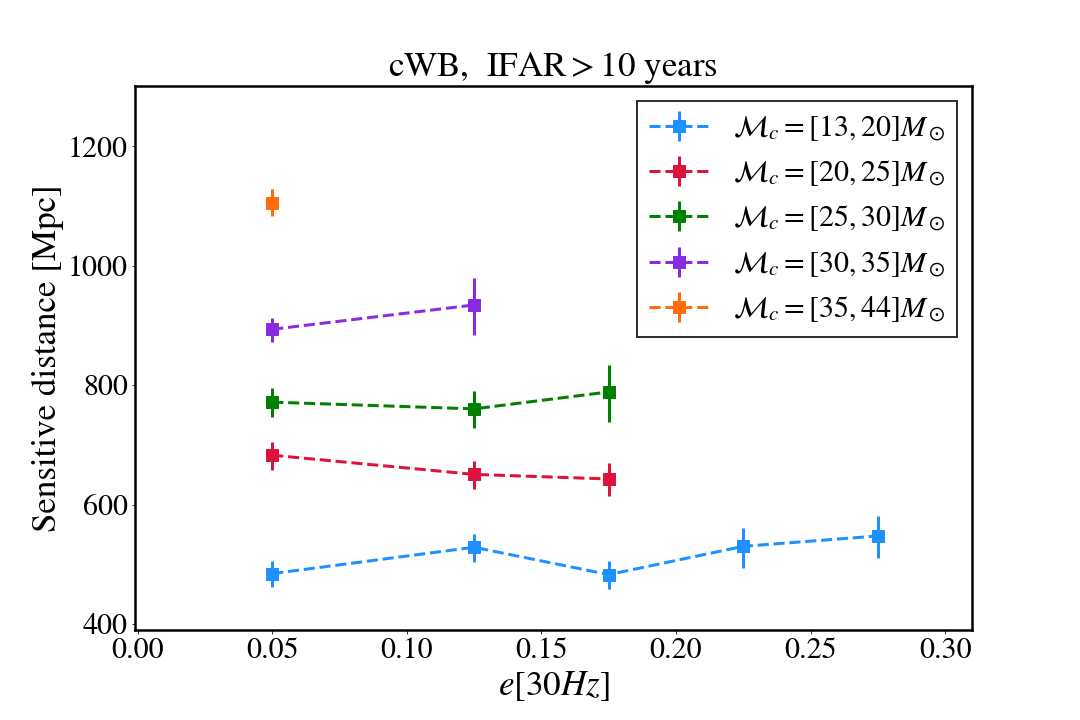}
    \includegraphics[scale=0.23]{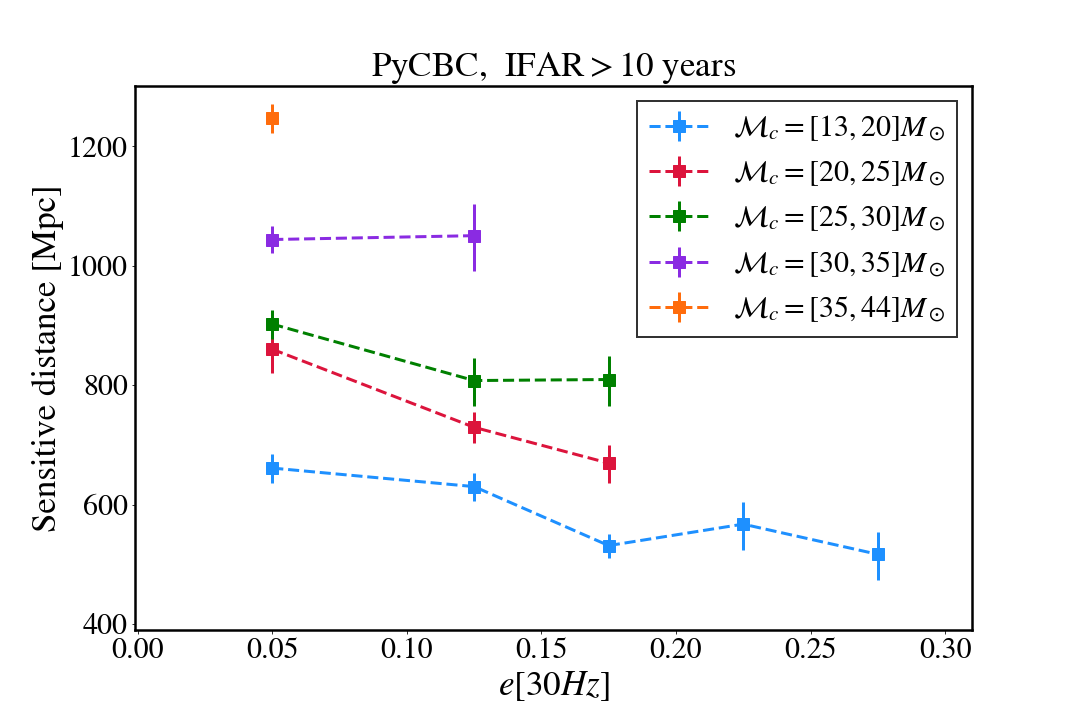}
   	\includegraphics[scale=0.23]{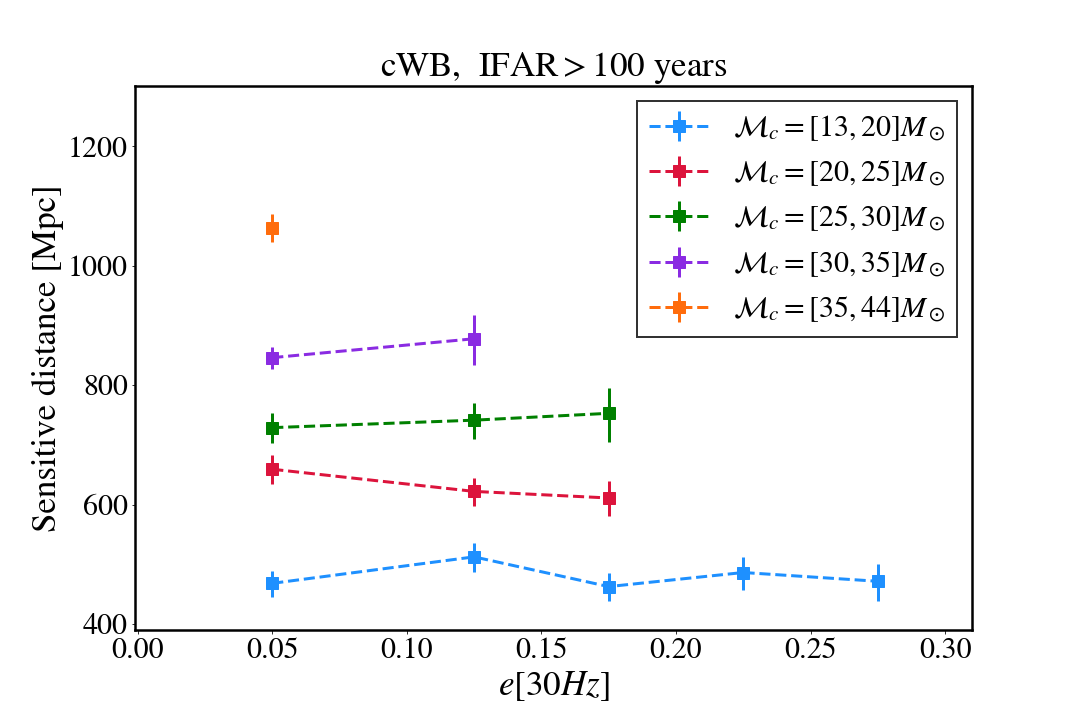}
    \includegraphics[scale=0.23]{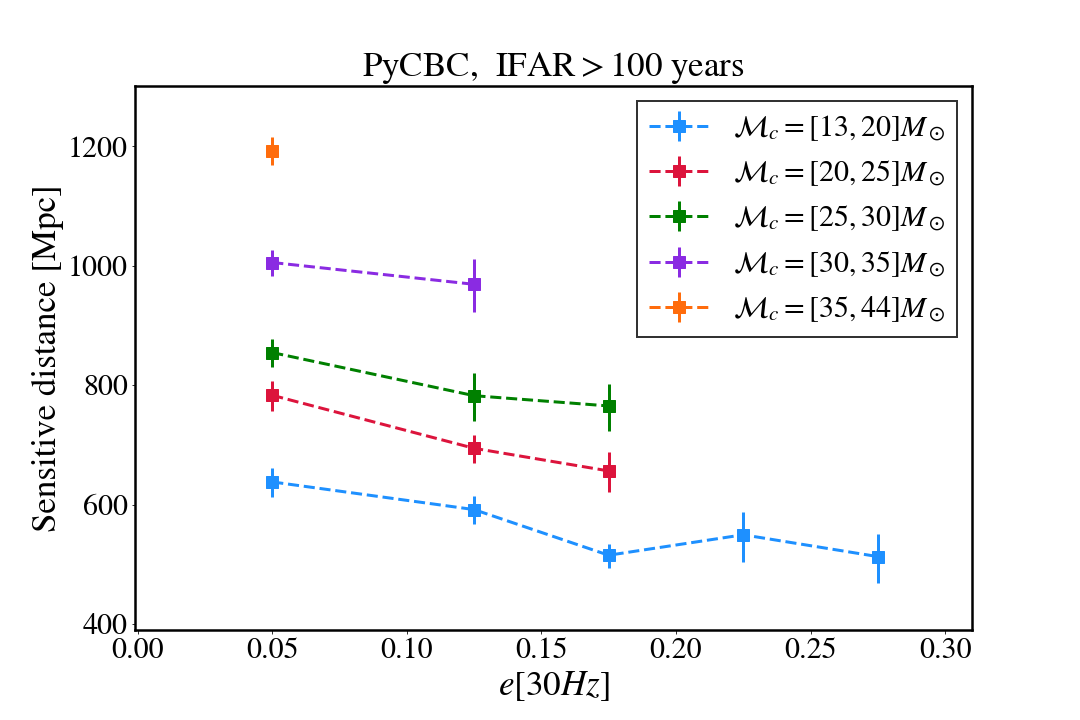}
\caption{Top row: The left and right panel show the sensitivity range for the cWB and PyCBC search pipelines for various chirp mass bins as a function of eccentricity defined at $30$Hz at an IFAR $>$ 10 years, respectively. Bottom row: The left and right panels display the sensitivity range for the cWB and PyCBC search pipeline for various chirp mass bins as a function of eccentricity defined at $30$Hz at an IFAR $>$ 100 years, respectively. The plot markers are placed in the center of the eccentricity bins.
}
\label{fig:range_chirp}
\end{figure*}

%\begin{figure}
%		\includegraphics[scale=0.23]{plots_paper_bo/plot13cwbIFAR10_01bin.png}
		%\includegraphics[scale=0.23]{plots_paper_bo/plot13pycbcIFAR10_01bin.png}
	%\caption{Sensitivity range for the PyCBC (cWB) search pipeline for various chirp mass bins as a function of eccentricity defined at $30 Hz$ at an iFAR $\rangle$ 10 years. The plot markers are placed in the center of the eccentricity bins.}
	%\label{fig:range_chirp}
%\end{figure}

\section{Search Sensitivity} \label{sec:introIFAR}

\subsection{Visible range}  \label{sec:introVolume}

The visible range for a given source parameters is calculated by injecting simulated waveforms into the data \cite{Usman:2015kfa,Nitz_2017, Klimenko:2008fu}. The False Alarm Rate (FAR) is a statistic measuring the frequency with which the search would rank non-astrophysical events with a detection statistic comparable to the one of a candidate event. In practice, each recovered signal is assigned an inverse false alarm rate  (IFAR=1/FAR) according to its detection statistic.  Then, one can compute for each bin of the source parameters the visible volume over a certain IFAR threshold. For a generic binary, the sensitive volume $V$ of a network of detectors with a given sensitivity can be defined as
\begin{equation} 
V(\xi) = \int_0^{\infty} f(z|\xi) \frac{\mathrm{d} V_c}{\mathrm{d}z} \frac{1}{1+z} \ dz,
\label{eq:eq10}
\end{equation}
where $f(z|\xi)$ is the detection probability of a binary with a given parameter set $\xi$ at redshift $z$, averaged over the extrinsic binary orientation parameters \cite{RnPo2}. In Eq. \eqref{eq:eq10} the sensitivity is assumed to be constant over the observing time, $T_\mathrm{obs}$, which is why we have chosen the specific chunk of O2 data where sensitivity was almost uniform. 

Given a population with parameters $\theta$, the total observed volume can be computed as 
\begin{equation}
    V_{\theta} = \int_{\xi} p(\xi | \theta) V(\xi) \ d\xi,
    \label{eq:eq11}
\end{equation}
where $p(\xi | \theta)$ describes the underlying distribution of the intrinsic parameters. The visible range can be then estimated as the radius of the visible volume.

The sensitivity of GW searches
% go for the searches here, since this is what we continue to talk about.
%\shubh{Indeed, I think maybe instead of saying GW searches one could say, The sensitivity of GW detectors towards Compact Binary Coalescence is a .. } 
is a strongly dependent function of the binary chirp mass and distance, and it also varies with spin. We also note that the eccentricity can be a relevant factor depending on the pipeline used to conduct the search. Thus, we have mainly chosen chirp mass binning to study the impact of eccentricity on visible range as it shows more clearly the dependence of the search sensitivity than other parameters, like the total mass.

\subsection{Injection set} \label{sec:InjSet}

The injection set used in this study is composed of the NR waveforms listed in Table \ref{tab:tabNRinj}. As a consequence, injections have fixed spin vectors and mass ratio values corresponding to those of the NR waveforms. Nonetheless, the total mass of the system acts as a scale parameter which can be freely specified, subject only to consistency with the length of the NR waveforms such that the injected signals start at the specified starting frequency in the band of the detectors. Due to the length limitations of the NR waveforms we set the starting frequency of the injection set to $30Hz$, which is also the frequency at which the value of eccentricity is specified.

We have chosen the number of performed injections to limit the computational cost necessary to run the search pipelines. We note that PyCBC is more expensive than cWB, however it also allows to achieve higher IFAR values. The largest subset of injected waveforms corresponds to $q=1$ nonspinning with 17317 injections distributed among cases with IDs 1, 2, 3 and 4 in Table \ref{tab:tabNRinj}. While the number of injections for the rest of waveforms has been decreased substantially due to the limited computational resources available to 3591 for $\chi_{\text{eff}}<0$ cases (IDs 5, 6 and 7), 4723 for $\chi_{\text{eff}}>0$ configurations (IDs 8, 9 and 10), 6416 for $q=2$ simulations  and 4458 for $q=4$ simulations (IDs 15, 16 and 17). As a consequence, the equal mass non-spinning eccentric case provides better statistics and permits to clearly identify the behavior of the sensitivity of both pipelines for specific values of chirp mass and eccentricity, as shown in Sec. \ref{sec:EffectEccentricity}.

The injection set is constructed using a uniform distribution in distance scaled by the chirp mass \cite{chirp_dist}. The total mass values are uniformly distributed from a minimum value consistent with the length of the NR waveforms, between $[30-50] M_\odot$ for our dataset, to a maximum total mass of $100 M_\odot$.

The orbital eccentricity of the individual injections, defined at a reference frequency of $30 Hz$ is estimated through Eq. \eqref{eq:eq000}. We note that with this method the maximum eccentricity at $30 Hz$ of a given injected NR waveform is given by the values of the last column of Table \ref{tab:tabNRinj}, as these values are measured at the start of the NR waveforms.

The moderate values of eccentricity considered here are well-suited for a first study of the sensitivity of gravitational waves searches to full IMR signals. Furthermore, many astrophysical models for eccentric binary black hole coalescences in the frequency band of ground-based detectors predict similar eccentricity values as those used here \cite{eastro1,eastro2,eastro3,PhysRevD.97.103014}.  

\subsection{Effect of eccentricity on search sensitivities}   \label{sec:EffectEccentricity}
We now turn to discussing  the visible range at IFAR thresholds of 10 and 100 years for both search pipelines and the same injection set. Although matched-filter searches are an optimal method to search for signals of known morphologies, in the case of eccentric BBHs computationally efficient waveform models describing the full GW signal of eccentric BBH coalescences have not yet been developed. For this reason it is expected that the quasicircular template bank used by PyCBC will not be able to detect eccentric BBH events with orbital eccentricities beyond a certain threshold. On the other hand, cWB does not require signal models for detection and thus its sensitivity to eccentric BBH signals is expected to only vary significantly as a function of signal strength, but only weakly in terms of other parameters like eccentricity.  It should be noted, however, that cWB is not an optimal method to detect BBH merger events and thus has lower sensitivity than PyCBC for regions of parameter space which are either explicitly covered by the PyCBC template bank or where the signal is otherwise `mimicked' by templates in the bank.

\begin{figure}
		\includegraphics[scale=0.25]{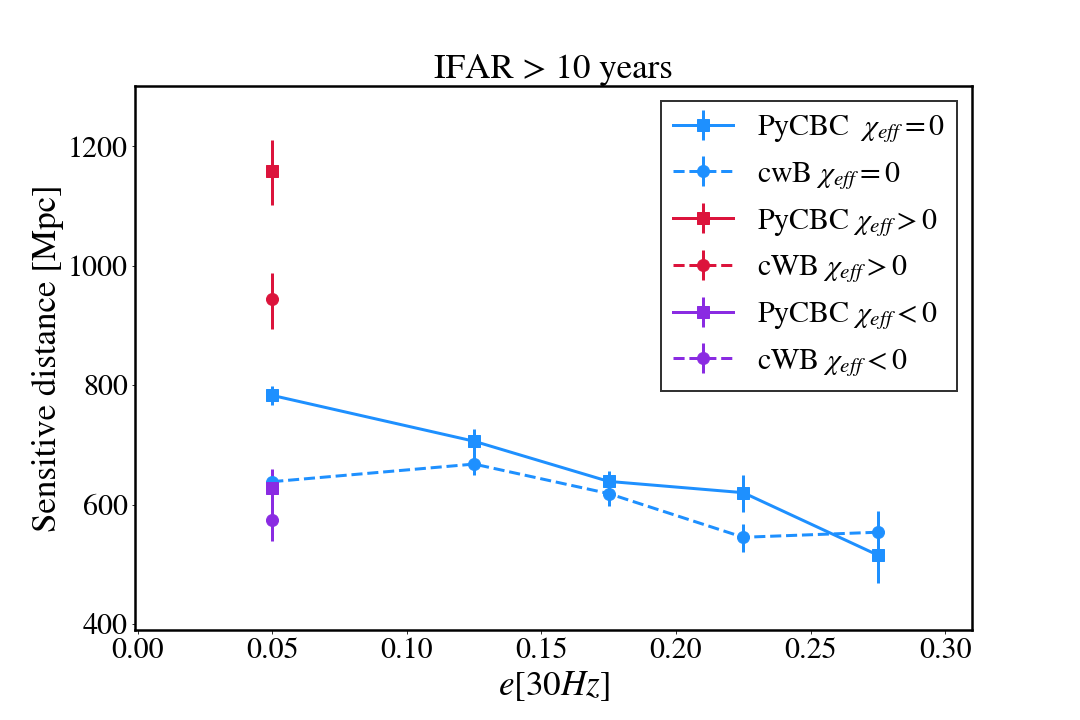}
		\includegraphics[scale=0.25]{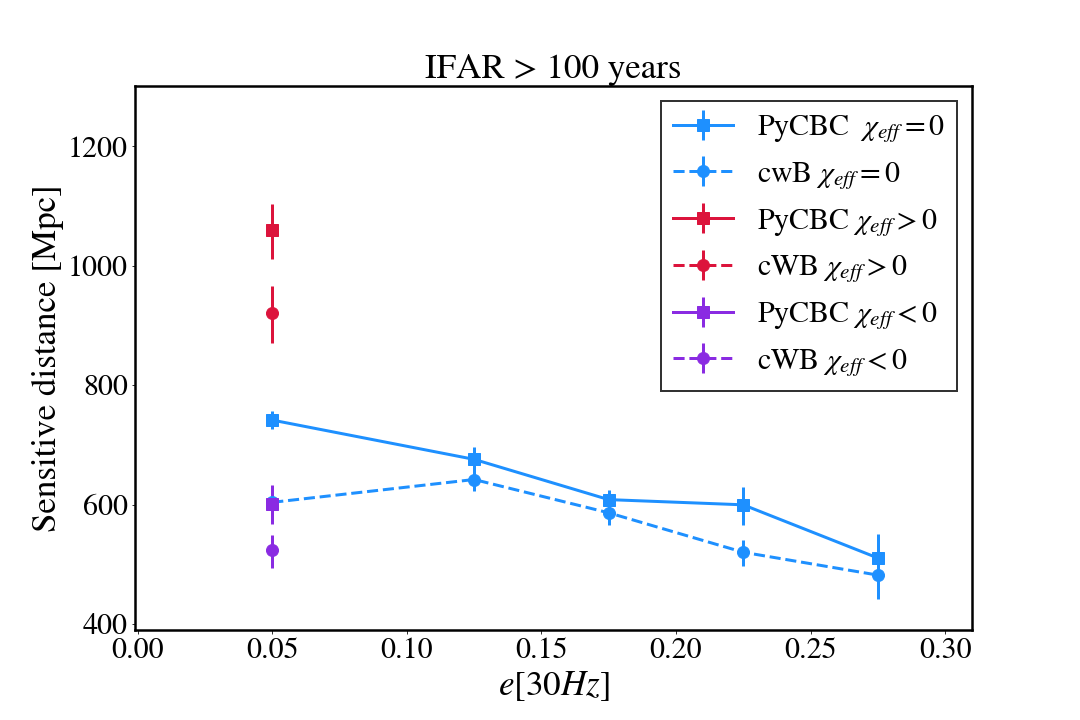}
	\caption{The upper (lower) panel shows the sensitivity range for the cWB and PyCBC search pipelines for equal mass non-spinning, positive spins and negative spins as a function of eccentricity defined at $30$Hz at an IFAR $>$ 10 years (IFAR $>$ 100 years). The plot markers are placed in the center of the eccentricity bins.}
	\label{fig:range_chieff}
\end{figure}

In Figure \ref{fig:range_chirp} we exhibit the visible ranges of the PyCBC and cWB pipelines binned in chirp mass and eccentricity for all the injected signals. The results show  a reduction in visible range of PyCBC with increasing eccentricity. The steepness of the reduction of visible range becomes more apparent when one goes to lower chirp masses; this is due to the fact that for high chirp masses the number of cycles visible in the sensitivity band of the LIGO detectors (and hence the inspiral part of the signal where eccentricity effects are pronounced) is rather short. One can conclude that for high chirp mass events with moderate to low eccentricities the PyCBC search and its quasicircular template bank does not lose much visible range. This behaviour is contrary to the low chirp mass case with moderate eccentricities, where the loss in visible range is substantial.

\begin{table}[h!]
\begin{center}
\resizebox{5.5cm}{!}{
 \def\arraystretch{1.3 }
\begin{tabular}{c c c c c c }
\hline  
\hline
ID &  $M_T [M_\odot]$ & $D_L [\text{Mpc}]$  &     $e_{\text{inj}} $ & $\mathcal{M}^{\text{Ecc}}$ & $\mathcal{M}^{\text{QC}}$ \\
%\hline
\hline
9 &46.4  &  277  &  0.11 &  0.95  &   0.86   \\
%\hline
12 & 96.4   & 200   & 0.04  &  0.98  &    0.93      \\
%\hline
\hline
\hline
\end{tabular}
}
\end{center}
\caption{Summary of the injected signals in Fig. \ref{fig:reconstructed_waveform}. The first column denotes the identifier of the simulation. The next columns indicate the total mass, $M_T$, the luminosity distance, $D_L$, the eccentricity, $e_{\text{inj}} $, of the injection, the match between the injected signal and the recovered one by cWB, $\mathcal{M}^{\text{Ecc}}$, and the match between the injected signal and the QC template with the same injected parameters, $\mathcal{M}^{\text{QC}}$.}
\label{tab:tabInjRecon}
\end{table}

Regarding cWB, previous work \cite{ebbho1o2} found that the search pipeline is almost independent of eccentricity for a given chirp mass bin. However, the waveforms used in that investigation \cite{East:2012xq} were based on geodesics in Kerr spacetime and the quadrupole formula for energy loss, and  significantly less accurate than the NR simulations used here.

We note an interesting feature in the dependency of the range as a function of eccentricity for cWB for the lowest chirp mass bin at IFAR$>$ 10 years. The range increases slightly as a function of eccentricity. This might probably be attributed to the power content in higher harmonics in eccentric BBH signals which is enhanced when the eccentricity increases. cWB captures the total excess power in the network of detectors and therefore can observe eccentric BBH events at larger range.  However, we note that this particular small increase in sensitivity is also compatible with a constant sensitive distance as the values are within the statistical error bars.
We find that our results are robust when changing the IFAR threshold from 10 to 100 years: sensitivity results for the higher IFAR choice are shown in the lower panels of Fig. \ref{fig:range_chirp}. One observes the expected overall decrease of the sensitive distance of both pipelines with increasing IFAR. Moreover, it can be noted that the dependence of the visible range on eccentricity retains the same features as at IFAR$>10$ years for both pipelines.

\begin{figure}
		\includegraphics[scale=0.35]{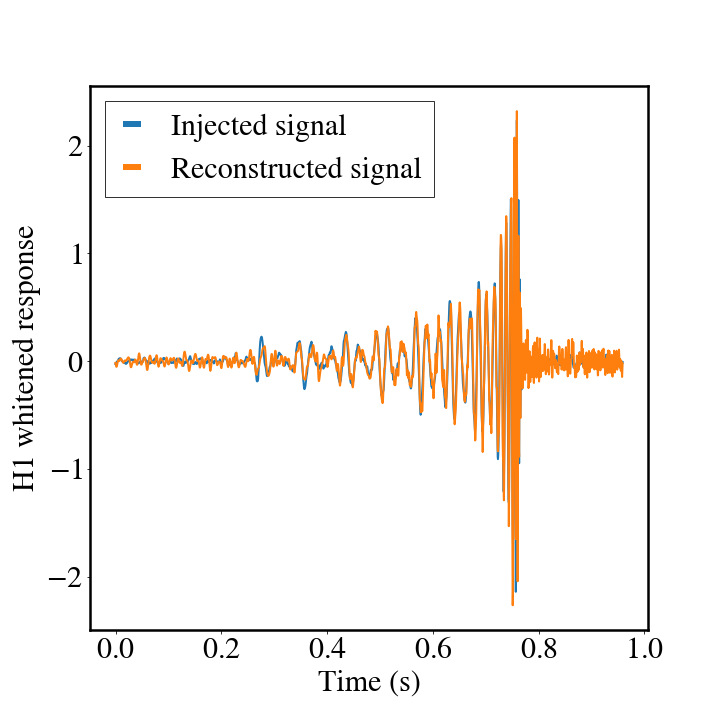}
		\includegraphics[scale=0.35]{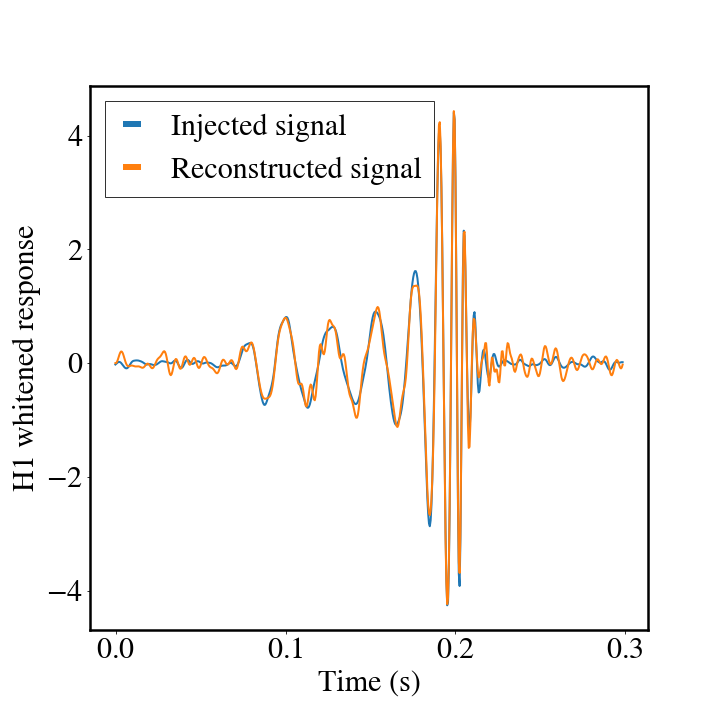}
	\caption{In the upper and lower panels the whitened detector response of LIGO Hanford, H1, in time domain are displayed for an injected signal corresponding to the waveform with ID 9 and 12 from Table \ref{tab:tabNRinj}, respectively.  Some of the parameters of the injected signals are shown in  Table \ref{tab:tabInjRecon}. In both panels the injected and reconstructed signals are represented with blue and orange colors, respectively.}
	\label{fig:reconstructed_waveform}
\end{figure}

In our NR simulations we only have waveforms with moderate eccentricities ($e_{30 Hz}$ $<$ 0.3). The subset of spinning waveforms is even more restricted in eccentricity values ($e_{30 Hz} < 0.12 $). As a consequence, a study of the impact of the effect of eccentricity and spins is more difficult. In Fig. \ref{fig:range_chieff} we show the sensitive distance for the equal mass spinning and non-spinning eccentric waveforms as a function of eccentricity for the chirp mass bin $[13,30]$ $M_\odot$. As expected, one observes that PyCBC has larger sensitivity for positive spins than for negative spins, as for positive spins  the matched filter pipeline can collect more SNR than for negative spins. There is also a drop in sensitivity for PyCBC with increasing eccentricity, while for cWB the small drop in sensitivity is consistent with statistical error bars. We note here that the results for the spinning waveforms are computed over a smaller number of injections compared to the nonspinning case, as explained in Sec. \ref{sec:InjSet}. In addition the small range in eccentricity of the spinning simulation does not allow to identify specific trends for the sensitivity as a function of the eccentricity. We point out that more insightful results could be obtained by increasing the number of injections and the range of values of initial eccentricity of the waveforms, and we leave the study of a large parameter space of the eccentric non-precessing spin sector, as well as eccentric spin-precessing systems, to future work.

%We note that the addition of spins to to mildly eccentric signals does not significantly change the detection efficiency for both pyCBC and cWB searches. We leave the study of a large parameter space of the eccentric non-precessing spin sector, as well as eccentric spin-precessing systems, to future work.

 Finally, we illustrate an example of the robust waveform reconstruction procedure of cWB \cite{Salemi:2019uea} applied to eccentric signals. In Fig.~\ref{fig:reconstructed_waveform} we display the whitened detector response of LIGO Hanford (H1) to two eccentric injected signals and the corresponding reconstructed waveforms by cWB, specifically for injections corresponding to the cases with ID 9 and 12 from Table \ref{tab:tabNRinj}, with injection parameters specified in Table \ref{tab:tabInjRecon}. We have also calculated the match, defined as the noise weighted inner product \cite{Jaranowski2012},  between the injected and recovered signal with cWB obtaining high agreement between both. The results of the match are also reported for a quasicircular template waveform computed against the same injected signals. One observes that the match against the quasicircular template, calculated using the SEOBNRv4 waveform model \cite{Bohe:2016gbl}, decreases significantly with increasing eccentricity, from $\mathcal{M}=0.93$ for ID 12 to  $\mathcal{M}=0.85$ for ID 9,  while the drop in the match against the cWB reconstructed waveform goes  from $\mathcal{M}=0.98$ for ID 12 to  $\mathcal{M}=0.95$ for ID 9. For cWB the match decreases because the signal with ID 9 is longer than for ID 12, and the reconstruction is expected to degrade the longer the signal is, while for the quasicircular template the increase of eccentricity decreases substantially the match due to the inability to resemble eccentric features in the injected signal. 
 
It should also be remarked that for cWB as an unmodelled search algorithm a high match between reconstructed and injected signal does not directly translate into having a high sensitivity of the pipeline as it is not a matched filter search pipeline. However, from a waveform modelling perspective it is still relevant to observe the ability of cWB to reconstruct the eccentric signal and the inability of the quasicircular template to resemble the injected signal with increasing eccentricity. As expected the reconstructed signal degrades after the waveform peak, thus, the ringdown is poorly reproduced due to the decrease in power of the signal. With these examples we want to illustrate the capability of cWB to recover features of eccentric signals, and we leave a thorough analysis of the reconstruction procedure of cWB applied to eccentric BBH signals for future work.

%We leave a thorough analysis of the reconstruction procedure of cWB applied to eccentric BBH signals for future work. \shubh{As can be seen in our illustration cWB is capable to recover the features of eccentric orbits,  a study of the reconstruction of eccentric BBH with cWB would be interesting. We leave a thorough analysis of the reconstruction capabilities of cWB applied to eccentric BBH signals for future work.}

%\toni{{\em{Moved from Appendix}} In Sec. \ref{sec:EffectEccentricity} we have discussed the effect of eccentricity in PyCBC and cWB at an  IFAR threshold of $10$ years. For completeness we present here results at an IFAR threshold of $100$ years. The outcome of such a calculation is shown in Fig. \ref{fig:range_chirp100}, where one observes comparing to Fig. \ref{fig:range_chirp} the expected overall decrease of the sensitivity distance of both pipelines with increasing IFAR. Moreover it can be noted that the dependence of the visible range as a function of eccentricity retains the same features for both pipelines.  }

\subsection{Comparisons of search sensitivities and astrophysical implications}     \label{sec:Comparisons}

\begin{figure}
		\includegraphics[scale=0.25]{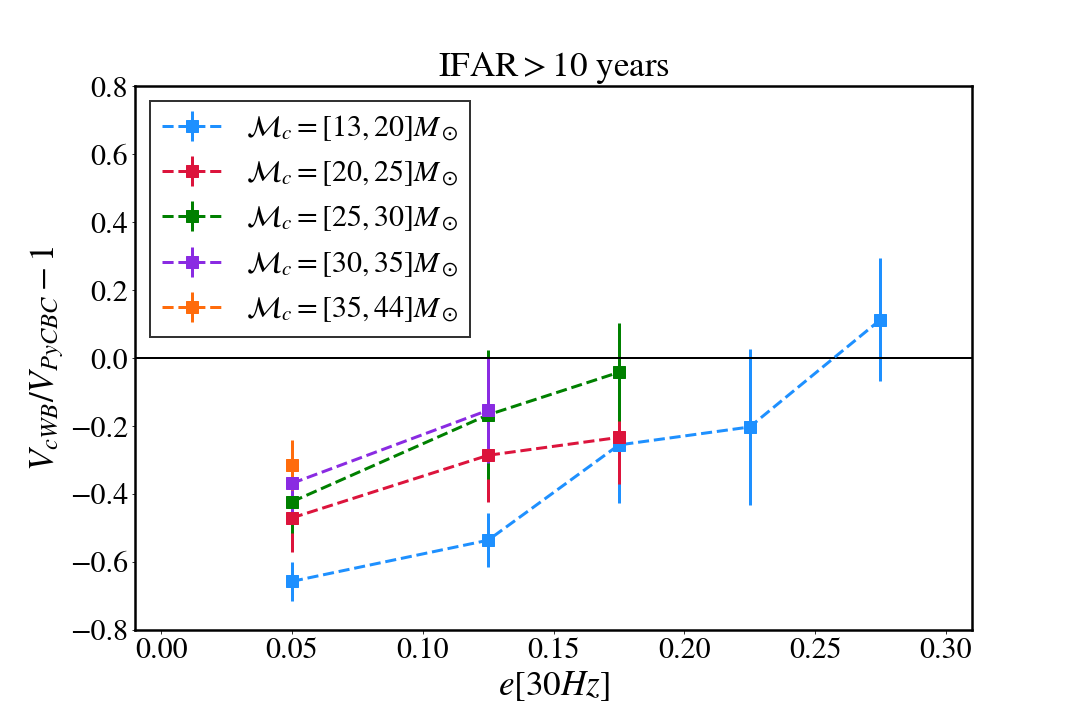}
		\includegraphics[scale=0.25]{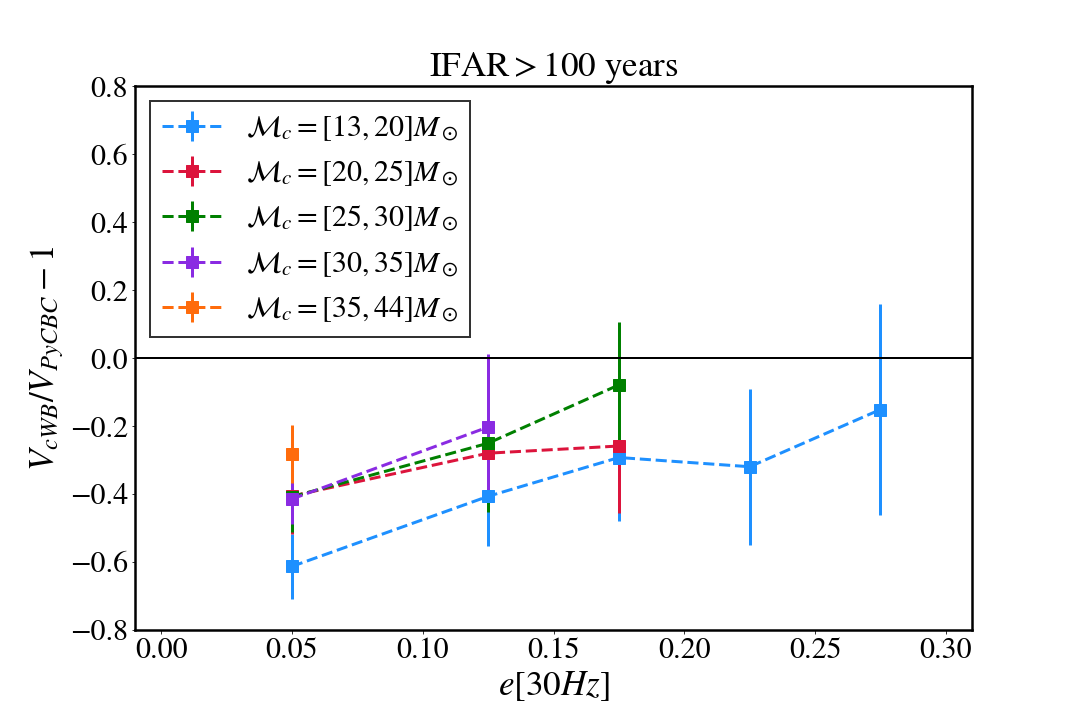}
	\caption{In the upper (lower) panel the relative difference in sensitivity volume between the search sensitivity of pyCBC and cWB for various chirp mass bins is presented as function of eccentricity at an IFAR $>$ 10 (IFAR $>$ 100) years. The plot markers are placed in the center of the eccentricity bins.}
	\label{fig:vol_comparison}
\end{figure}

In Figure \ref{fig:vol_comparison} we show the comparison of the visible volumes of PyCBC and cWB at IFAR thresholds of 10 and 100 years. Within our injection set PyCBC almost always performs better than or similar to cWB in terms of visible volume. In the case of low chirp mass and high eccentricity the situation is reversed: PyCBC loses sensitivity and cWB becomes more sensitive, specially at  IFAR $>$ 10 years.

In the lower panel of Fig.~\ref{fig:vol_comparison}  the comparison in sensitive volume between both pipelines at IFAR $>$100 years is shown. One observes that the trends in relative volume between both pipelines are similar to the case at an IFAR threshold of 10 years, although the error estimates are larger due to the decrease in sensitivity of both pipelines with increasing IFAR values.  The results show that PyCBC has a larger sensitive volume than cWB, even for the low chirp mass bin. However, with increasing eccentricity the decrease in sensitive volume of PyCBC and the constancy of the cWB one, make the relative volume to be an increasing function, which for higher values of eccentricity is expected to cross zero, as it is the case at IFAR $>$10 years.

These comparison results can also be viewed in the light of coalescence rate. Suppose the coalescence rate of eccentric BBH mergers with eccentricities between $(0,0.3)$ at 30 Hz is $R_{eBBH}$; then the number of visible events will be simply $N_{events} = R_{eBBH} \times V_{iFAR} \times T_{obs}$. The relative difference in the number of detected events will be the same as the relative difference between the visible volume for the two search algorithms that we have considered. From this, we can conclude that at IFAR$>10$ years cWB  will see $\sim$10 \% more events than PyCBC if the chirp mass is between  $[13 M_{\odot}, 20 M_{\odot}]$ and the eccentricity at 30 Hz is between [0.25-0.3].  

%\toni{We also show in Fig. \ref{fig:vol_comparison100} the comparison in sensitive volume between both pipelines at IFAR $>$100 years. The results show that PyCBC has a larger sensitive volume than cWB even for the low chirp mass bin. However, with increasing eccentricity the decrease in sensitive volume of PyCBC and the constancy of the cWB one, make the relative volume to be an increasing function, which for higher values of eccentricity is expected to cross zero, as it is the case at IFAR $>$10 years of Fig. \ref{fig:vol_comparison}. Thus, this shows the complementarity between both pipelines to detect eccentric binaries. For moderate eccentricities $(e<0.3)$ PyCBC outperforms cWB, although with increasing eccentricity tables turn, as it is shown for IFAR$>10$ years in Fig. \ref{fig:vol_comparison}.}

%\toni{Moved from Appendix. The sensitive volumes of  PyCBC and cWB at an IFAR threshold of $10$ years have been compared in  Sec.  \ref{sec:Comparisons}. }

%Eccentricity decays very rapidly in bound BBH systems, retaining very high eccentricity till very late times in the evolution of BBH systems is not very frequent. However, there are models which predict high eccentricities in the sensitivity band of LIGO detectors \cite{highE}. \\ 

\section{Conclusions} \label{sec:summary}

In this paper we have quantified for the first time the sensitivities of GW search algorithms to eccentric BBH signals, using NR simulations of eccentric BBH mergers. The effect of eccentricity on matched filtered searches has  only been studied for inspiral-only waveforms until now \cite{brown}; we have extended those studies to complete IMR signals. The search range of unmodeled searches for eccentric signals has been previously investigated with a particular IMR waveform model  \cite{ebbho1o2}; however, that waveform model is far less accurate than the NR simulations used here.

We have employed two different gravitational wave searches for BBHs to compare the search sensitivity in terms of visible volume. The matched filter search PyCBC performs better than the unmodeled search cWB in most parts of the limited parameter space that we have considered. Only in the parameter space region of low chirp mass and high eccentricity does cWB perform better than PyCBC. It should also be noted that the parameter space that is covered by our NR injections is rather small. Due to the restricted length of the NR simulations, the parameter space of low chirp mass ($\mathcal{M}_c < $ 13) and high eccentricity $e_{30 Hz} > 0.3 $ is not yet probed in this work. This, however, is the most interesting part of parameter space for eccentric BBHs, with waveform morphologies that are substantially different than those of quasicircular BBHs. We plan to investigate this part of parameter space in subsequent work, with eccentric hybrid waveforms that combine NR data with an analytic description of the inspiral, or with future waveform models for the full IMR signal. 

The two search pipelines used here \textemdash very different algorithms as described in Sec. \ref{sec:introSearches} \textemdash offer a complementary way to search for BBH mergers in different parts of the source parameter space. Constructing a template-based search for eccentric BBH will be challenging as the rate of background triggers increases with the increase of template bank parameters. In the light of astrophysical considerations \cite{O_Leary_2006,PhysRevD.97.103014}, most of the BBH events observable by LIGO and Virgo are expected to have eccentricities lower than $0.2$ at 30 Hz; this region of parameter space has been demonstrated to be well-covered by the PyCBC search, even with a quasi-circular template bank. Certain astrophysical scenarios suggest LIGO-Virgo relevant BBH events with higher eccentricities: for such sources the cWB search provides decent coverage.  

With the  expected availability of computationally efficient and accurate eccentric IMR BBH waveforms models (and/or eccentric hybrids) in the near future it will be interesting to probe the low chirp mass and high eccentricity part of the parameter space, where the modelled search is penalized due to substantial dephasing between the quasicircular template bank and the signal. 

With future upgrades \cite{Aasi:2013wya}, the detectors' low-frequency sensitivity (in the range $24-100$ Hz) is expected to improve significantly; this will in turn allow a significant gain in SNR during the inspiral even for BBH systems with relatively high masses, adding more prominence to detectable inspiral features like eccentricity and penalizing the matched-filter searches for eccentric BBH even further. With future improvements at low frequencies, the role of un-modeled searches is therefore expected to become important also for the part of parameter space which is well-covered by matched-filter searches at current detector configuration.

\section{Acknowledgements}  \label{sec:Acknowledgements}
We would like to thank Debnandini Mukherjee for useful comments about the manuscript. This work was supported by the Spanish Ministry of Education and Professional Formation grants FPU15/03344.  The author also acknowledges the support by the Govern de les Illes Balears through the Vicepresidència i Conselleria d’Innovació, Recerca i Turisme and the Direcció General de Política Universitària i Recerca with funds from the Tourist Stay Tax Law  ITS 2017-006 (PRD2018/24), the European Union FEDER funds and EU COST Actions CA18108, CA17137, CA16214, and CA16104, the Ministry of Science, Innovation and Universities and the Spanish Agencia Estatal de Investigación grants FPA2016-76821-P, RED2018-102661-T, RED2018-102573-E, FPA2017-90566-REDC, FPA2017-90687-REDC, and the Generalitat Valenciana (PROMETEO/2019/071). The author thankfully acknowledges the computer resources at MareNostrum and the technical support provided by Barcelona Supercomputing Center (BSC) through Grants No. AECT-2020-1-0025, AECT-2019-3-0020, AECT-2019-2-0010, AECT-2019-1-0022, AECT-2018-3-0017, AECT-2018-2-0022, AECT-2018-1-0009, AECT-2017-3-0013, AECT-2017-2-0017, AECT2017-1-0017, AECT-2016-3-0014, AECT2016-2-0009, from the Red Española de Supercomputaci{\'o}n (RES) and PRACE (Grant No. 2015133131). {\tt ET} simulations were carried out on the BSC MareNostrum computer center under PRACE and RES allocations and on the FONER cluster at the University of the Balearic Islands. The authors are grateful for computational resources provided by the LIGO Laboratory and supported by National Science Foundation Grants PHY-0757058 and PHY-0823459. MH acknowledges support from Swiss National Science Foundation (SNSF) grant IZCOZ0-177057. ARB is grateful to the Pauli Center for Theoretical Studies at ETHZ which provided valuable travel support during stages of this work. S.T. is supported by Forschungskredit Nr. FK-19-114 and Swiss National Science Foundation grant number 200020 182047. This research has made use of data, software and/or web tools obtained from the Gravitational Wave Open Science Center (https://www.gw-openscience.org), a service of LIGO Laboratory, the LIGO Scientific Collaboration and the Virgo Collaboration. LIGO is funded by the U.S. National Science Foundation. Virgo is funded by the French Centre National de Recherche Scientifique (CNRS), the Italian Istituto Nazionale della Fisica Nucleare (INFN) and the Dutch Nikhef, with contributions by Polish and Hungarian institutes. The authors gratefully acknowledge the support of the NSF CIT cluster for the provision of computational resources for pyCBC and cWB runs. 

\appendix

\bibliography{main}
\bibliographystyle{unsrt}

\end{document}